\date{}
\documentclass[10pt,aps,prl,twocolumn,superscriptaddress]{revtex4-1}
\usepackage{graphicx,psfrag,amsmath,amssymb,amsfonts,color}
\usepackage[T1]{fontenc}
\usepackage{natbib}
\usepackage{hyperref}

\newcommand{\ket}[1]{\ensuremath{\,|{#1}\rangle}}

\newcommand{\beq}{\begin{equation}}
\newcommand{\eeq}{\end{equation}}
\newcommand{\bea}{\begin{eqnarray}}
\newcommand{\eea}{\end{eqnarray}}

\begin{document}
\title{Optical super-resolution sensing of a trapped ion's wave packet size}

\author{Mart\' in Drechsler}
\affiliation{Departamento de F\'isica, FCEyN, UBA and IFIBA, UBA CONICET, Pabell\'on 1, Ciudad Universitaria, 1428 Buenos Aires, Argentina}
\affiliation{QUANTUM, Institut f\"ur Physik, Universit\"at Mainz, Staudingerweg 7, 55128 Mainz, Germany}

\author{Sebastian Wolf}
\affiliation{QUANTUM, Institut f\"ur Physik, Universit\"at Mainz, Staudingerweg 7, 55128 Mainz, Germany}

\author{Christian T. Schmiegelow }
\affiliation{Departamento de F\'isica, FCEyN, UBA and IFIBA, UBA CONICET, Pabell\'on 1, Ciudad Universitaria, 1428 Buenos Aires, Argentina}

\author{Ferdinand Schmidt-Kaler}
\affiliation{QUANTUM, Institut f\"ur Physik, Universit\"at Mainz, Staudingerweg 7, 55128 Mainz, Germany}

\begin{abstract}
We demonstrate super-resolution optical sensing of the size of the wave packet of a single trapped ion. Our method extends the well known ground state depletion (GSD) technique to the coherent regime.
Here, we use a hollow beam to strongly saturate a coherently driven dipole-forbidden transition around a sub-diffraction limited area at its center and observe state dependent fluorescence.
By spatially scanning this laser beam over a single trapped $^{40}\mathrm{Ca}^+$ ion, we are able to measure the wave packet sizes of cooled ions.
Using a depletion beam waist of $4.2(1)\,\mu$m we reach a spatial resolution which allows us to determine a wave packet size of $39(9)\,$nm for a near ground state cooled ion.
This value matches an independently deduced value of $32(2)\,$nm, calculated from resolved sideband spectroscopy measurements. Finally, we discuss the ultimate resolution limits of our adapted GSD imaging technique in view of applications to direct quantum wave packet imaging.

\end{abstract} 

\maketitle
Super-resolution fluorescence microscopy is an impressive group of well established techniques in optical imaging, capable of producing sub-diffraction spatial resolution while maintaining the advantages of optical microscopy. The first known technique, the so-called stimulated emission depletion microscopy (STED), was proposed and demonstrated by S.~Hell~\cite{hell1994breaking, Klar:99}. There, a sample containing fluorescent marker molecules is illuminated by a light field combining a Gaussian excitation beam with a hollow-structured depletion beam~\cite{hell1994breaking}. The fluorescence (exciting) beam is counteracted by the structured depletion beam in such a way that fluorescence can only be emitted inside its dark center. This way, a sub-diffraction sized effective point spread function (ePSF) arises from the combined action of the fluorescence and depletion beams. Scanning both beams over the sample and detecting the emitted fluorescence results in an imaging resolution which may reach $< 50\,$nm~\cite{hein2008stimulated,persson2011fluorescence} in case of biological samples and down to $2.4\,$nm for color centers in diamond~\cite{wildanger2012solid}. 
Also, there exist approaches that propose using coherence in optical transitions to produce a resolution enhancement~\cite{gerhardt2010coherent,zeng2015sub}, as shown in quantum dots~\cite{kaldewey2018far}.
Here we demonstrate a related method which resolves the motional wave packet of single trapped atoms cooled close to their ground state.

Confinement and cooling of single atoms and ions provide wave packets that are localized at a scale well below $100\,$nm and are highly controllable. Therefore a super-resolution technique would be necessary to image and determine properties of these systems. Pioneering experiments employed light fields inside a cavity to trace a single atom trajectory~\cite{hood1998real,HORAK2002} or inversely, a single trapped ion was used as a localized probe to sense the field inside an optical cavity~\cite{guthohrlein2001single,mundt2002coupling}. Imaging of these systems is done by collecting emitted fluorescence with high numerical aperture (NA) objectives which result in diffraction limited images. Nevertheless, analyzing the signal with statistical methods makes it possible to estimate the center of the wave packet with a precision of a few nanometers~\cite{wong2016high,alberti2016super,araneda2019wavelength} or to observe micromotion~\cite{PhysRevA.103.023105}. Moreover, the positioning of ions and atoms~\cite{schmiegelow2016phase} in optical standing-wave fields allows for sharply resolving its nodes or imaging atomic density distributions ~\cite{mcdonald2019superresolution,subhankar2019nanoscale, yang2018theory}.
In this work, we pave the way towards direct imaging of the motional wave function of a single ion.

\begin{figure}[ht]
\centering
\includegraphics[width=0.5\textwidth]{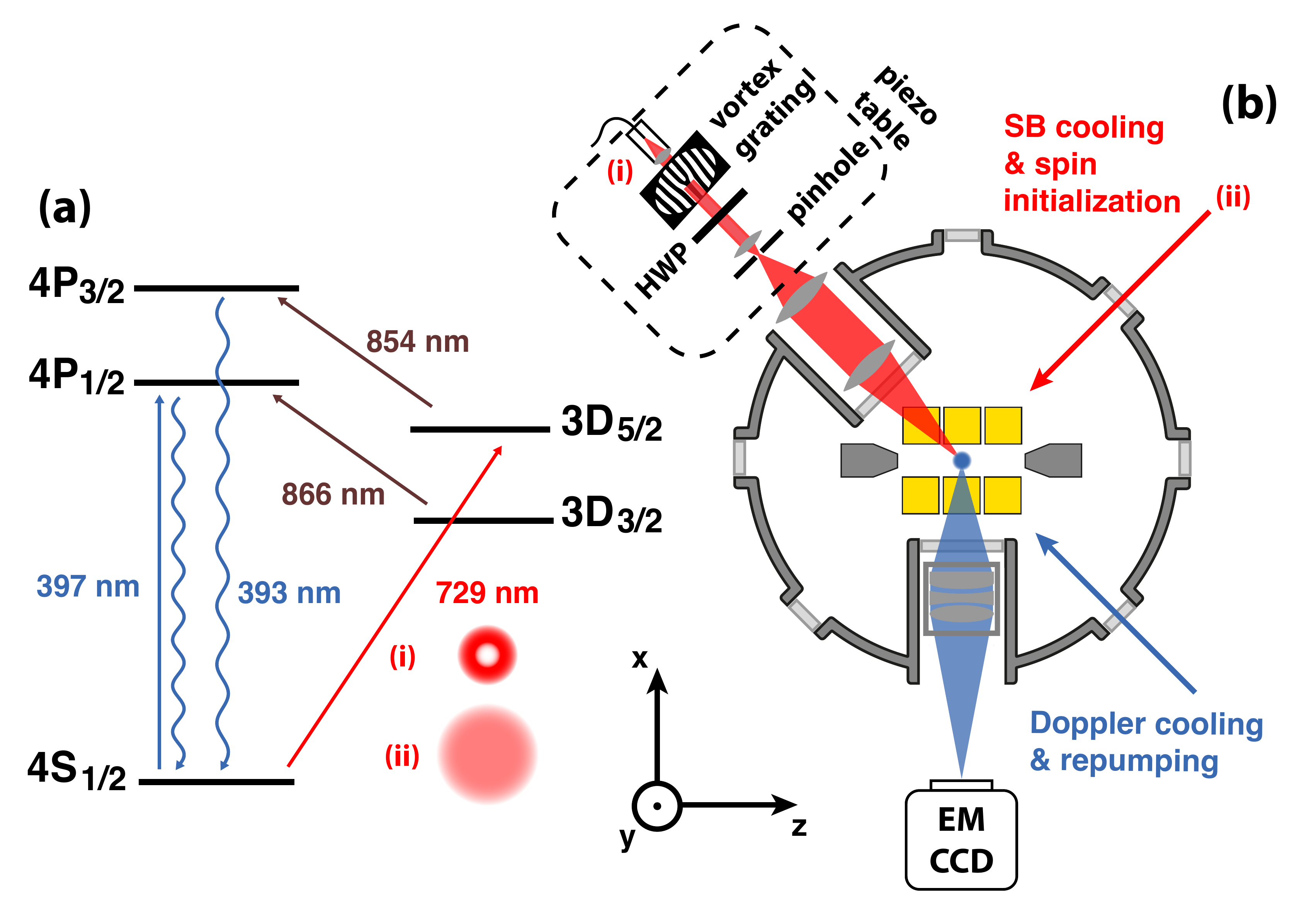}
\caption{(a) Energy levels and transitions in $^{40}$Ca$^+$:  Excitation on the S$_{1/2}$ - P$_{1/2}$ dipole transition near $397\,$nm is used for Doppler cooling, and we observe emitted fluorescence decay of the P$_{1/2}$ state for state detection. The dipole-forbidden S$_{1/2}$ - D$_{5/2}$ transition near $729\,$nm is used as a vortex beam (i) to probe the spatial extent of the ion's wave packet. Using a regular Gaussian beam (ii) we perform resolved sideband-cooling and -spectroscopy to determine the mean phonon number.  (b) Sketch of the setup: The ion is trapped in a segmented linear ion trap (yellow), the ion's fluorescence light (blue) is collected by an $f/1.6$ objective and imaged onto an EMCCD. The vortex beam (red, i) shape is created from the output of a single mode fiber via a holographic diffraction grating and focused to the ion with a $f=75\,$mm lens ($\text{NA}=0.2$). To scan the vortex beam waist over the ion wave packet a 3D Piezo translation stage is used. A second beam (red, ii) near $729\,$nm is used for resolved sideband cooling. The coordinates showed here allow to define all the relevant frames of references: for the trap $(\sqrt{2}\hat{x}_t, \sqrt{2}\hat{y}_t, \hat{z}_t) = (\hat{x}+\hat{y}, \hat{y}-\hat{x}, \hat{z})$ and for the beam $(\sqrt{2}\hat{x}_B, \hat{y}_B, \sqrt{2}\hat{z}_B) = (\hat{x}+\hat{z}, \hat{y}, \hat{z}-\hat{x})$ respectively.}
\label{fig:scheme}
\end{figure}

Inspired by STED and its closely related variant, ground state depletion microscopy (GSD)~\cite{PhysRevLett.98.218103}, we implement an optical super-resolution method for sensing the wave packet size of a single ion.  To implement our proposal we choose an ion species with a short-lived transition used for fluorescence and a long-lived non-bleachable transition used for depletion. The latter is driven coherently and allows for the high saturation needed to achieve sub-diffraction resolution. 
In our setup, trapping occurs independently of the optical field, thus allowing for positioning and manipulation of the atomic wave packet freely. Using optical cooling (Doppler or sideband) we can prepare different wave packet sizes in the range of tens of nanometers and then measure them with the proposed method.

We demonstrate our method by imaging a trapped $^{40}$Ca$^+$ ion, with energy levels and transitions as in Fig.~\ref{fig:scheme}a). Laser-induced fluorescence is obtained on the S$_{1/2}$~-~P$_{1/2}$ dipole transition near $397\,$nm. Since the ion may decay into the metastable 3$D_{3/2}$ level, laser light near $866\,$nm must also be used to continuously repump the population out of the 3D$_{3/2}$ level back into the S-P cycle. For GSD, the S$_{1/2}$ ground state is depleted into the metastable 3D$_{5/2}$ level using a dipole-forbidden transition near $729\,$nm. The decay time of the metastable D-level ($\approx1.2\,$s) exceeds the dipole decay time from the P-level by more than 8 orders of magnitude.  This allows us to use these two transitions as depletion and fluorescence, correspondingly. For the depletion transition, the light-ion interaction is described by the Jaynes-Cummings model: Laser radiation resonant to the depletion transition switched on during an interaction time $\tau$, leads to coherent Rabi oscillations between populations of both involved states S$_{1/2}$ and D$_{5/2}$, respectively. The Rabi frequency $\Omega(I)=\sqrt{  I(r)~(3\lambda^3\Gamma_\text{D})/(4\pi^2 \hbar c)}$ depends on the intensity $I(r)$ at the ion position, wavelength $\lambda$ of the laser and the decay rate $\Gamma_\text{D}$ of the D state. As a depletion laser, we use a first-order Laguerre-Gauss beam tuned to the S-D transition near $729\,$nm. As the beam is hollow-shaped, transitions to the 3D$_{5/2}$ state will be induced, unless the ion is at the dark center of the beam. 

We estimate the size of this dark area, corresponding to the point spread function (ePSF) of our sensing technique, as a function of the depletion beam power $P_0$, the interaction time $\tau$ and the hollow-beam waist size $w_0$. In the vicinity of the dark center with $r\ll w_0$, the intensity distribution of a first order Laguerre-Gauss beam scales as $I(r) \sim  4 r^2 P_0/(\pi w_0^4)$, which in turn leads to an expression for the spatial-dependent Rabi frequency $\Omega(r)$. Note that this approximation is ensured  by the vortex topology of the beam, and is robust against slight optical misalignment or lens errors. Also, we consider the intensity profile constant along the beam's direction, since the Rayleigh range is much bigger than the ion's spatial extension. For ground-state cooled ions the depletion probability reads $P_\text{D}(r) = \sin^2(\tau \Omega(r)/2) \approx (\tau \Omega(r)/2)^2$, approximated for the small argument of the $\sin^2$ function. As we are interested in the first oscillations, the sinusoidal approximation holds well even for non ground-state cooled ions in the  Lamb-Dicke limit. That is, as long as the wavelength of the excitation laser exceeds the ion's wave packet size, which is largely fulfilled for our case. This way, we can obtain the resolution $\sigma_\text{ePSF}$ as the standard deviation of the depletion probability $P_\text{D}(r)$:
\begin{align}
\sigma_\text{ePSF} = \sqrt{\frac{\pi^3 \hbar c}{3 \lambda^3 \Gamma_\text{D}}} \, \frac{w_0^2}{\tau \sqrt{P_0}} \sim 75 \,\text{nm} \frac{w_0^2 [\mu \text{m}]}{\tau[\mu \text{s}] \sqrt{P_0 [\text{mW}]}}.
\label{eq:Dpsf}
\end{align}

One can see that $\sigma_\text{ePSF}$ is proportional to the square of the waist size $w_0^2$, but reduces with increasing laser power $P_0$ and increasing laser interaction time $\tau$. We note that the resolution scales as $\tau^{-1}$ with the pulse time. This is a strong advantage with respect to incoherent situations in STED or GSD microscopy. With a modest waist of $5\,\mu$m, a laser power of $1\,$mW and a pulse length of $20\,\mu$s, $\sigma_\text{ePSF}$ becomes already $90\,$nm. We will discuss the refined theory of this scaling in Tab.~\ref{table:errors} and the Supplemental Material~\cite{onlinematerials}. 

\begin{figure*}[htp]
\centering
\includegraphics[width=0.95\textwidth]{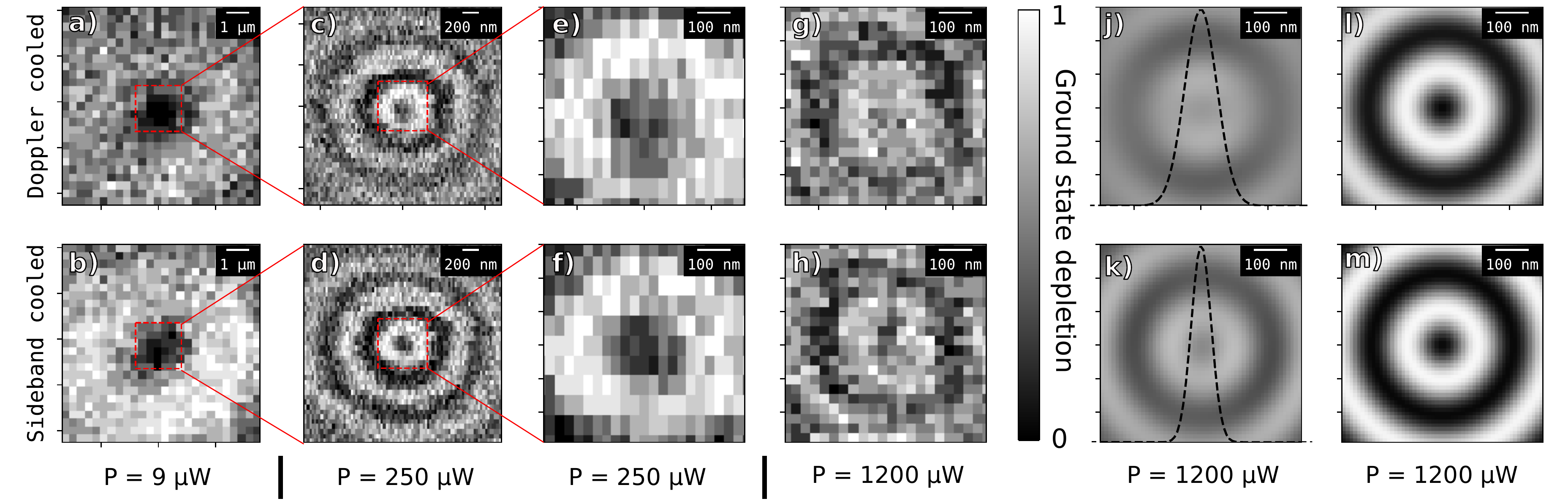}
\caption{\textbf{Measured and simulated GSD profiles.} The gray-scale indicates the probability of depleting the ground state with the hollow beam pulse. Results for Doppler cooling (top row) and for sideband ground state cooling (bottom row) are displayed. Each pixel GSD value is determined as the average of 10 repetitions of a measurement sequence. The statistics follow the expected quantum projection noise. (a, b) Large area scans at low excitation power $9\,\mu$W, exhibiting the beam shape. (c, d) At increased power of $250\,\mu$W Rabi oscillations become visible. (e, f) Zooming in at $250\,\mu$W power. (g, h) Increasing  power to $1200\,\mu$W. Here, the ePSF is sufficiently small to discriminate the size of the wave packet: for Doppler cooling, center contrast is lost, while narrow wave packet from ground state cooling preserves contrast at center. (l, m): Simulations of ePSF assuming a point-like particle with Rabi oscillations corresponding for Doppler- and ground state cooling, respectively. (j, k)  ePSF convoluted with spatial wave packet (dashed curve) for Doppler- and ground state cooling of the axial mode, respectively. Note, that loss of contrast at the center allows us to distinguish both cases and matches with data in (e, f).}
\label{fig:images}
\end{figure*}

We carry out experiments on a single ion held in a linear segmented Paul trap with harmonic confinement at frequencies $\omega_{\{i\}}$ with $\{i = x_t,\,y_t,\,z_t\}$ inside an ultra high vacuum with view-ports for laser beam access and fluorescence light collection, sketched in Fig.~\ref{fig:scheme}b). Electrodes for $(x_t,y_t)$-confinement are separated by $960\,\mu$m and operated at a frequency of $2\pi\times30\,$MHz with an voltage amplitude of $100\,\text{V}_\text{pp}$. This results in radial frequencies of $\omega_{rad} \sim 2\pi\times 1.5\,$MHz for $^{40}$Ca$^+$. The axial frequency is controlled by DC-voltages on the segments and on the endcaps~\cite{wolfthesis}, $\omega_z = 2\pi\times760\,$kHz under typical operation conditions. 

We can control the wave packet size by two parameters: the confinement frequencies $\omega_{i}$, which are set by the trapping parameters and the average phonon occupation numbers of the harmonic oscillator states $\bar{n}_i$, which are determined by the laser cooling processes. The wave packet size for the harmonic oscillator ground state reads as $\sigma_{i}^0=\sqrt{\hbar/m \omega_{i}}$. For the tightly confining potential in radial $(x_t,y_t)$-directions we calculate $\sigma^0_{rad}= 13\,$nm, and in axial direction $\sigma^0_{z} = 18\,$nm, respectively. For thermal states, the wave packet remains of Gaussian shape, with increased width $\sigma_{i} = \sigma^0_{i} \sqrt{2\overline{n}_{i}+1}$. 
For an ion at the Doppler cooling limit we have in the axial $z$-direction,  $\overline{n}_z \simeq 10$ which amounts to a wave packet size of $\simeq 80\,$nm, in the stronger confined radial directions $\overline{n}_{rad} \simeq 5$ corresponding to $\simeq 45\,$nm. Additionally we apply resolved sideband cooling  \cite{PhysRevLett.83.4713} on the axial harmonic oscillator and reach $\overline{n}_z=1.1(2)$, which corresponds to a wave packet size of $32(2)\,$nm. This value is determined from resolved sideband Rabi oscillations, see Supplemental Material~\cite{onlinematerials}. This way, we have an independent, indirect method of determining the wave packet size in $z$-direction, which reduces from the Doppler cooled size of $\simeq 80\,$nm, down to $32(2)\,$nm for the sideband cooled case.

We obtain images, as shown in Fig. \ref{fig:images}, by performing two-dimensional scans by moving the beam in the $\hat{y} = \hat{y}_B$ direction and the ion in the $\hat{z} = \hat{z}_t$ direction. For each position we execute the following sequence: i) Preparation of the internal electronic and the external motional state of the ion by Doppler cooling, and optionally sideband cooling of the axial mode, followed by optical pumping into the $m=-1/2$ level of the S$_{1/2}$ ground state (see Supplemental Material~\cite{onlinematerials} for details on the spin initialization scheme). ii) A depletion Laguerre-Gaussian beam, generated by a grating-hologram and focused to $4.2(1)\,\mu$m waist, is applied for $\tau = 19\,\mu$s. The beam's waist was measured from Fig.~\ref{fig:images}a). It has orbital angular momentum $l=-1$ and carries circular polarization $\sigma=-1$. A magnetic field of $0.25\,$mT is aligned in direction with the $\vec{k}_{729}$ vector. We tune the laser frequency resonant to the $\ket{4^2\text{S}_{1/2}, m=-1/2} \leftrightarrow \ket{3^2\text{D}_{5/2}, m=-3/2}$ transition. iii) The sequence is concluded by  fluorescence readout: we switch on laser radiation near $397\,$nm and $866\,$nm and record the ion's fluorescence during a $3\,$ms interval on the EMCCD. From the photon statistics and the lifetime of the D state we reveal a detection fidelity better than $97.5\%$. 
Repeating the sequence (i) to (iii) and averaging, yields a GSD value of this image pixel. One single sequence execution takes typically $5-10\,$ms. Including  communication time, a 20 $\times$ 20 pixel image with 10 repetitions per pixel is recorded in roughly 2~minutes, see Fig.~\ref{fig:images}. Note that the times used for initialization and fluorescence detection are by far exceeding those of the GSD laser pulse. Using a lens with a higher collection efficiency, a photomultiplier detector and optimizing communications could reduce the sequence time by about one order of magnitude.  

Our data shows the expected spatial resolution $\sigma_\text{ePSF}$ improvement, when increasing the power of the depletion vortex beam, in agreement with Eq.~\ref{eq:Dpsf}.  At low beam power, the peak intensity does not generate oscillations and the images resemble the intensity profile of the depletion beam, see a) and b) in  Fig.~\ref{fig:images}. When the power is increased, see c) to f), the GSD image shows an improved spatial resolution. The contrast in Rabi oscillations in the outer fringes is slightly reduced for the case of the Doppler cooled ion c) as compared to the sideband cooled ion d). This difference stems from the fact that the thermal distribution over different phonon eigenstates $|n\rangle$ leads to an averaging over a  distribution of different corresponding Rabi frequencies~\cite{wineland_review}. When the depletion power is further increased to yield $\sigma_\text{ePSF} \leq \sigma_\text{Doppler}$, the size of the wave packet under Doppler cooling, we observe a loss of visibility in the GSD at the center, see g). However, for the identical power the GSD contrast at the center remains visible, see h), if the ion is cooled to a narrower wave packet size. Here, and for a power of $1.2\,$mW, the spatial resolution is in the order of $\sim 30~$nm, improved by a factor of $\times12$ as compared to the non-saturated case, see b).

We compare our experimental results with our theoretical model by convoluting the expected ePSF and the the wave packet. To do so, we take into account spectroscopically determined axial and radial frequencies, the projections along the vortex beam direction and the full Jaynes-Cummings interaction at the estimated ion's temperature, see j) and k) of Fig.~\ref{fig:images}. As in the experimental data, we see a strong loss of contrast at the center in case of the Doppler cooled situation j) where the wave packet is $\sim 80\,$nm compared to the sideband cooled case k), where the wave packet is $\sim 30\,$nm.

It is important to note that the point spread function (ePSF) does not depend strongly on the average phonon number, as can be seen by comparing the simulated ePSFs in l) and m) of Fig.~\ref{fig:images}. There it can also be noted that the center feature does not depend on the ion's temperature. Therefore, the reduction in contrast measured at the center is an effect of the wave packet size and not of the temperature.

\begin{figure}[htp]
\centering
\includegraphics[width=0.45\textwidth]{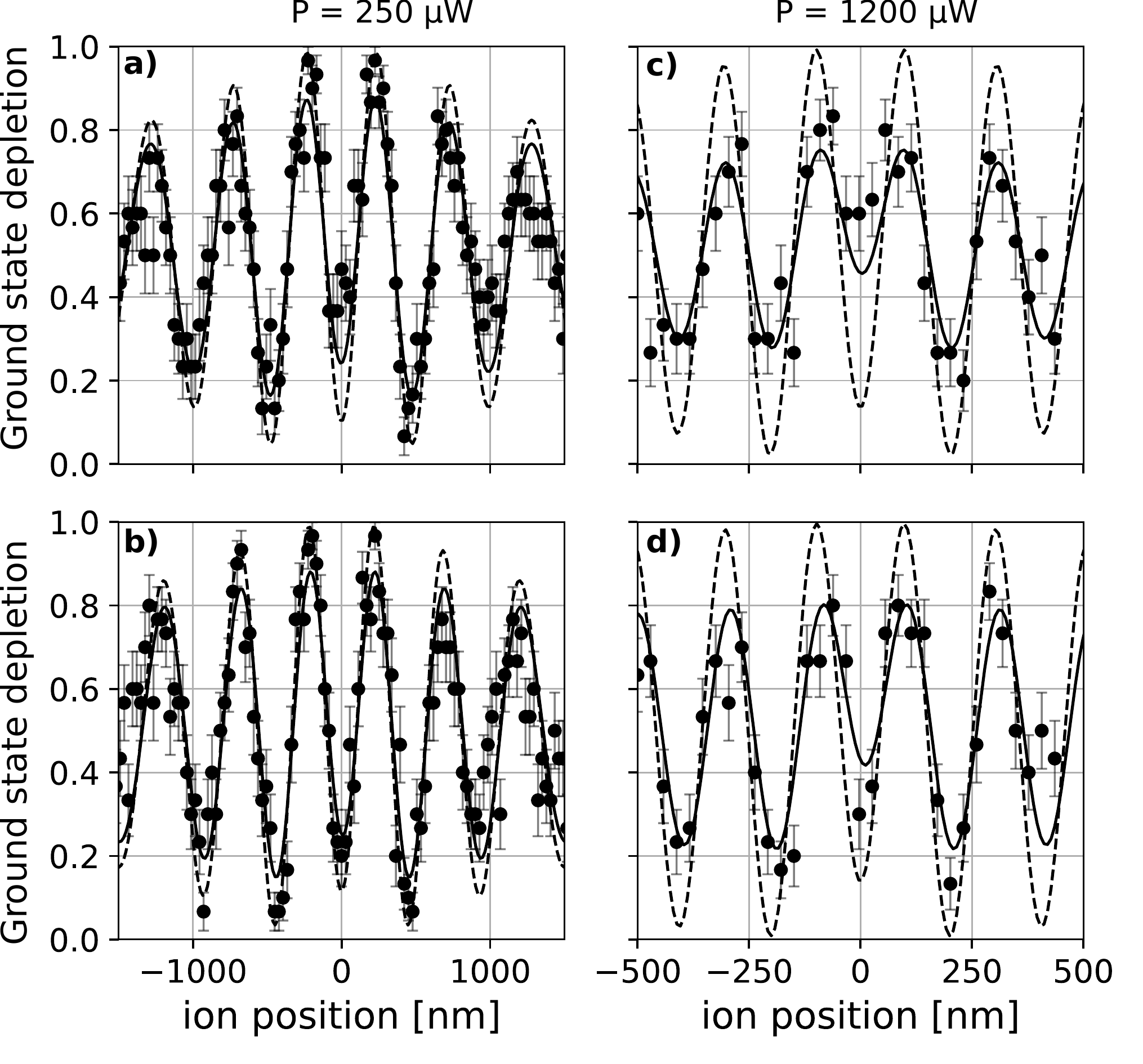}
\caption{\textbf{One dimensional GSD profiles} by averaging the three middle rows of data (solid circles) from Fig.~\ref{fig:images} c,d,g,h (a,b,c,d here) with non linear fits with the full model (solid) and ePSF without wave packet convolution (dashed line). For a pulse power achieving a resolving $\sigma_\text{ePSF}$ similar to the wave packet size, both models strongly differ, and the fit allows for the extraction of a value for the wave packet size: the fit for the Doppler cooled ion outputs $\sigma_\text{z} = 60(5)\,$nm (c) and the one from the axially sideband cooled ion outputs $\sigma_z = 39(9)\,$nm (d).}
\label{fig:cuts}
\end{figure}

For a quantitative analysis we use a horizontal cut of the GSD image data, see Fig.~\ref{fig:cuts}. At medium depletion power, see a) and b), the $\sigma_\text{ePSF}$ still exceeds the wave packet size $\sigma$ such that only slight differences are observed between Doppler cooling and axial sideband cooling. For high power, see c) and d),  the wave packet size exceeds the $\sigma_\text{ePSF}$ and the GSD contrast is reduced at the center in different amounts depending on the cooling mechanism used. We fit a model function which takes into account the convolution with the projection of the wave packet size on the imaging plane (see Supplemental Material~\cite{onlinematerials} for details). We fix the radial phonon numbers to the Doppler limit $\overline{n}_\text{rad}=5$, since this number is the same for all experiments and allows the axial size and beam power to vary.
The fit reveals beam powers consistent with the values used in the experiment and, more interestingly, an axial wave packet size of $60(5)\,$nm for the Doppler cooled case and of $39(9)\,$nm, for the axially sideband cooled case, which matches the independently determined size of $32(2)\,$nm. 

Several light-matter interaction features can limit the achievable resolution of our method by making the depletion laser have non-zero-excitation probability at the center of the beam, altering its "dark" center. (i) An imperfect alignment of the magnetic field with respect to the beam's propagation direction would break the symmetry of the configuration changing selection rules~\cite{schmiegelow2012light}. (ii) An imperfect left-circular polarization would generate a longitudinal field at the center~\cite{quinteiro2017}. Also, spurious off resonant excitations could occur either because (iii) frequency components of the finite square pulse exist or due to (iv) power broadening. As a scaling non-dimensional quantity we introduce the saturation value $S\equiv P/P_{NS}$, where $P$ is the used power and $P_{\mathrm{NS}}$ is the maximum power that makes no resolution enhancement, which roughly produces the condition seen in Figs.~\ref{fig:images} a), b). In Table~\ref{table:errors} we show the limits on the achievable saturation values $S^\mathrm{lim}$ for our setup, obtained by constraining the spurious excitation probability to 0.01. All limits are well above the maximum values used in the presented measurements $S_{exp}\sim 1 - 130$. We also calculate the achievable resolution limit $\sigma_{ePSF}^{\mathrm{lim}}$ by inserting  $P_\mathrm{lim}=S_\mathrm{lim}P_\mathrm{NS}$ in Eq.~\ref{eq:Dpsf}, and find that it depends only on how well the different sources of spurious excitation can be suppressed, and not on the size of the beam waist. For more details on this, see section III of the Supplemental Material~\cite{onlinematerials}.

\begin{table}
\begin{ruledtabular}
\begin{tabular}{ccc}
Error type  & $S_{\mathrm{lim}}$ & $\sigma_\text{ePSF}^{\mathrm{lim}}$ \\\hline

$\vec{B},\vec{k}$ angle & $ 10^{2}(w_0 k)^2$ &  $ 9\,$nm \\ 
Pulse width & $10^3(w_0 k)^2$ & $ 4\,$nm \\  
Power broadening & $ 10^{4} (w_0 k)^2 $ & $ 1\,$nm \\ 
Polarization         & $10^{6}(w_0 k)^2$ & $ 0.1\,$nm 
\end{tabular}
\end{ruledtabular}
\caption{Estimation of parasitic excitation of the depletion transition,  scaling with the saturation parameter $S$ and limiting the achievable resolution. The last column shows the achievable limit under current operation conditions. These limits would be reached with higher powers than the used in the presented measurements.}
\label{table:errors}
\end{table}

In this work we have demonstrated the direct sensing of Gaussian-shaped spatial wave packets of a single ion at low temperatures by implementing super-resolution coherent GSD sensing. In this way we were able to directly reveal motional wave packet sizes of 39(9)~nm. Our experimental and theoretical investigation of current limitations shows the way to push the spatial resolution limit down to the range of a few nm. At that level, we plan for a sensing of non-classical wave packets of a single ion, such as the Fock states $n=1$ and $n=2$ or squeezed states of motion.

\begin{acknowledgments}
We acknowledge the support of the Alexander von Humboldt-Stiftung and the Deutscher
Akademischer Austauschdienst (DAAD), M.D. and C.T.S. acknowledge the support of ANPCyT, UBACyT, CONICET, and S.W. and F.S.-K. of the German Research Foundation within TRR 306 and by the Deutsche Forschungsgemeinschaft (DFG, German Research
Foundation) – Project-ID 429529648 – TRR 306 QuCoLiMa
(“Quantum Cooperativity of Light and Matter”).  We also acknowledge early-stage contributions from Laurence Pruvost and Elias Alstead, and Fransisco Balzarotti for comments on the manuscript.  

\textit{Note added.}---During the submission of the manuscript we became aware of the related work by
Qian et al \cite{qian2021nanosecond}. 
\end{acknowledgments}

\bibliography{ions}  

\end{document}


\title{Supplementary information: Optical super-resolution sensing of a trapped ion's wave packet size}

\author{Mart\' in Drechsler}
\affiliation{Departamento de F\'isica, FCEyN, UBA and IFIBA, UBA CONICET, Pabell\'on 1, Ciudad Universitaria, 1428 Buenos Aires, Argentina}
\affiliation{QUANTUM, Institut f\"ur Physik, Universit\"at Mainz, Staudingerweg 7, 55128 Mainz, Germany}

\author{Sebastian Wolf}
\affiliation{QUANTUM, Institut f\"ur Physik, Universit\"at Mainz, Staudingerweg 7, 55128 Mainz, Germany}

\author{Christian T. Schmiegelow }
\affiliation{Departamento de F\'isica, FCEyN, UBA and IFIBA, UBA CONICET, Pabell\'on 1, Ciudad Universitaria, 1428 Buenos Aires, Argentina}

\author{Ferdinand Schmidt-Kaler}
\affiliation{QUANTUM, Institut f\"ur Physik, Universit\"at Mainz, Staudingerweg 7, 55128 Mainz, Germany}

\maketitle

\widetext

\section{Spectroscopical determination of ion's temperature}

On the S-D cuadrupole transition it is possible to observe Rabi oscillations on the carrier but also on the sidebands, which gives useful methods for mean phonon number $\bar{n}$ measurement. One of the simplest and most robust method consist in the comparison of the probability $P_D(t)$ to end up in the D state after excitation of the ion on the red and blue sidebands when they are resonantly excited. In the Lamb-Dicke regime and assuming that the motional state after cooling has a thermal distribution, one can obtain the following expression~\cite{RevModPhys.75.281}:

\begin{align}
    P_D^{rsb}(t) = \frac{\bar{n}}{\bar{n+1}}P_D^{bsb}(t)
\end{align}

\begin{figure}[h]
\centering
\includegraphics[width=0.5\textwidth]{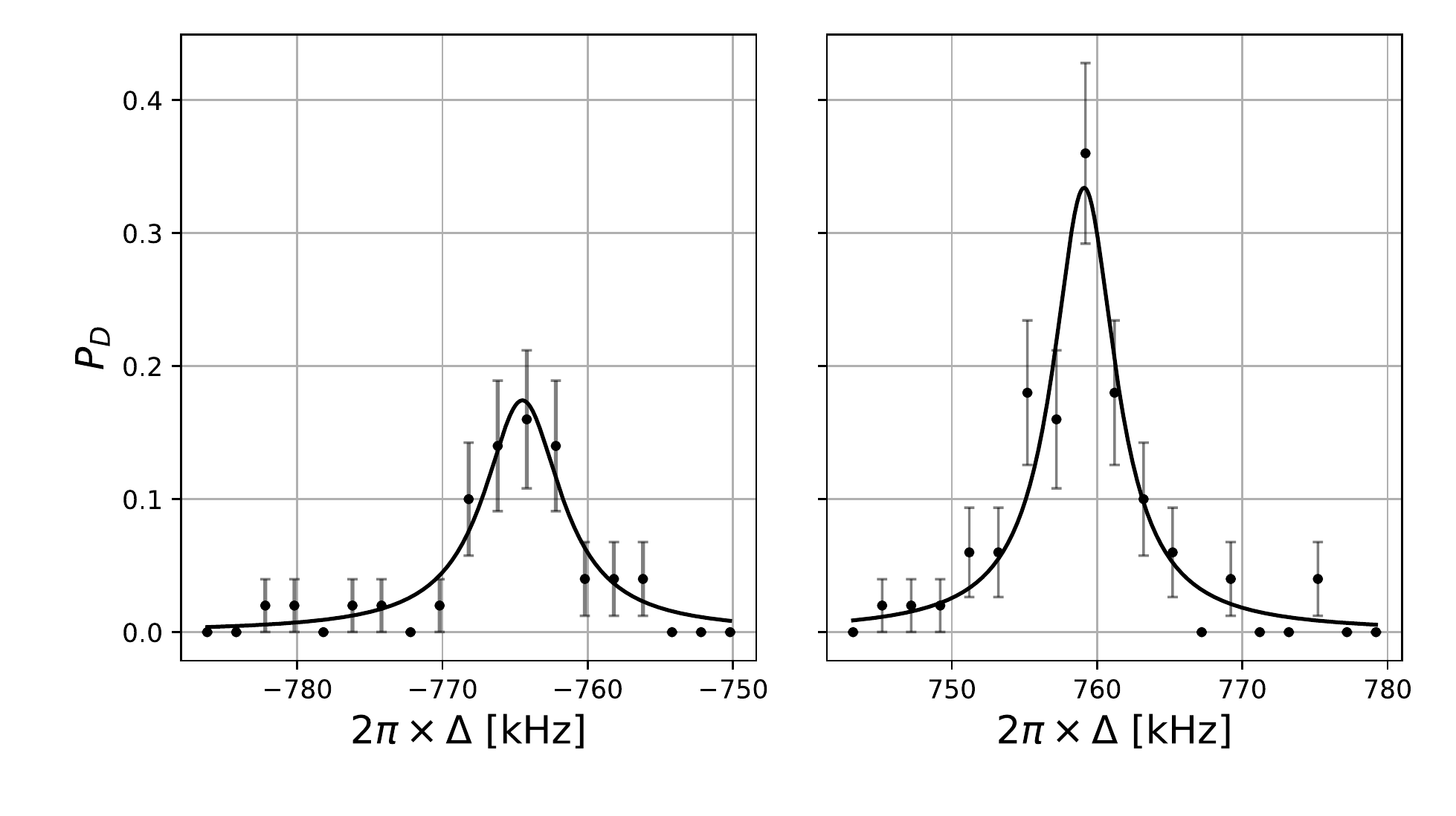}
\caption{\textbf{Sideband spectroscopy for $\bar{n}$ measurement.} The figure shows measurements of the first axial side-bands spectra (black dots), with a lorentzian fit (solid line) for each case. From the fit, we extract the excitation probabilities when the excitation beam is resonant with the sidebands, and from these values the axial mean phonon number gives $\bar{n}_{ax} = 1.1 \pm 0.2$. Cooling on the axial mode was done by applying 40 pulses on the 2nd red side-band, followed by 30 pulses on the 1st red side-band. The measurement sequence for each point of the spectra consisted on applying pulses of $100~\mu$s and repeating the experiment 50 times.}
\label{fig:sbs}
\end{figure}

Therefore the mean phonon number $\bar{n}$ is given by ratio of these probabilities $p = P_D^{rsb}/P_D^{bsb}$ as $\bar{n} = p/(1-p)$. In figure \ref{fig:sbs} we show a measured spectrum of both side-bands, using non saturating power and a pulse length of $100~\mu$s. 

\section{Optical pumping scheme}

In order to initialize the internal state of the ion into the  $m=-1/2$ level of the S$_{1/2}$ ground state, we implement the following sequence: a $\pi$-pulse on the $ \ket{4S^2_{1/2}, m = 1/2} \leftrightarrow \ket{ 3D^2_{5/2}, m = -3/2}$ transition is applied, which depopulates the unwanted $\ket{ 4S^2_{1/2}, m = 1/2}$ state. This is done with the SB cooling and Spin initialization laser beam near $729~$nm, see Fig. 1 of the main text. Then, a pulse on the $ 3D^2_{5/2} \leftrightarrow 4P^2_{3/2}$ dipole transition near $854~$nm resets the ion's state to the $4S^2_{1/2}$ manifold via spontaneous decay from $4P^2_{3/2}$. Repeating this sequence by 10 times, the state $\ket{ 4S^2_{1/2}, m = -1/2}$ is prepared at a fidelity $\gtrsim 99\%$.

\section{Light-mater interaction limiting resolution}

Several light-matter interaction features can limit the achievable resolution of our method by making the depletion laser have non-zero-excitation probability at the center of the beam, and therefore altering its "dark" center. (i) An imperfect alignment of the magnetic field with respect to the beam's propagation direction would break the symmetry of the configuration changing selection rules~\cite{schmiegelow2012light}. This would create excitations on the depletion transition due to the strong transverse field gradient at the center of the beam. Furthermore, and even though the depletion beam is tuned to the $\Delta m =-1$ Zeeman transition of the S-D manifold, other sub-levels could be excited off-resonantly. (ii) An imperfect left-circular polarization would generate a longitudinal field, which could excite the $\Delta m = 0$ transition, at the center of the beam~\cite{quinteiro2017}. Additionally, the transition with $\Delta m=-2$  has a non-zero excitation probability, also at the center of the beam. This could generate spurious off resonant excitation either because (iii) of the extended frequency components of the finite square pulse of the depletion beam or (iv) due to power broadening of the transition.

We consider here these four main contributions and quantify them. To do this, we calculate the spurious excitation probability p as a function of $S\equiv P/P_{NS}$, which is the ratio of the power used P with respect to the maximum power $P_{NS}$ that makes no resolution enhancement. At a power of $P_{NS}$, the Rabi frequency where the beam's intensity is maximum is $\Omega_0 = \pi/\tau$, since by definition we are asking that $ \sin^2(\Omega_0 \tau /2) = 1$. Therefore, we can also write $S$  as a ratio of Rabi frequencies $S = \Omega^2 /\Omega_0^2 =\Omega^2\tau^2/\pi^2$.

i) To turn off transitions with $\Delta m =-1$, created by the electric gradient , we must align the $\vec{k}_{729}$ and the total magnetic field on the ion $\vec{B}$ parallel. This alignment is estimated to be done with an error of $\theta_B < 3^\circ$, where $\theta_B$ is the angle between $\vec{B}$ and $\vec{k}_{729}$. Following \cite{schmiegelow2012light} one can calculate the ratio of excitation strengths as a function of this angle:

\begin{align}
    \frac{\Omega_{\vec{B}}}{\Omega} = \frac{\cos{\gamma}+2\cos{\gamma}\cos{\theta_B}+\sin{\gamma}}{\cos{\gamma}\cos{2\theta_B}+\cos{\theta_B}\sin{\gamma}} \, \sin{\theta_B}\,\frac{1}{\sqrt{2}\,k w_0} < 10^{-1} \times \frac{1}{\sqrt{2}\,k w_0} \
\end{align}

In this formula $\gamma$ depends on the beam polarization and was measured to be $3\pi/4 \pm 0.02$ (that is a $1^{\circ}$ error), which accounts for a left circularly polarized beam. $\Omega_{\vec{B}}$ and $\Omega$ are the Rabi frequencies of the spurious and depletion transitions, respectively. We can write both of them as $\Omega_{\vec{B}} = \Omega_0 \sqrt{S_{\vec{B}}}$ and $\Omega = \Omega_0 \sqrt{S}$. This way we can calculate how saturated this spurious transition is with respect to the depletion one:

\begin{align}
 \frac{S_{\vec{B}}}{S} =  10^{-2} \times \frac{1}{2\,(k w_0)^2}
\end{align}

Now, the unwanted excitation in the center is given by 
\begin{align}
      p_{\vec{B}} = \sin^2(\Omega_{\vec{B}} \tau/2) = \sin^2(\sqrt{S_{\vec{B}}}~ \pi/2) \sim (\sqrt{S_{\vec{B}}}~\pi/2)^2=\frac{\pi^2 }{8\,(k w_0)^2}~10^{-2}~S
\end{align}

 ii) Next, imperfection in the left circular polarization (we estimate an error of $1\%$ given by the polarization analyzer used) generates a longitudinal field at the center of the beam~\cite{quinteiro2017}. This can generate off-resonant transitions with $\Delta m = 0$, on the  $\ket{4^2\text{S}_{1/2}, m=-1/2} \leftrightarrow \ket{3^2\text{D}_{5/2}, m=-1/2}$ transition with $\Delta m=-2$ at $2\pi\times4\,$MHz detuning. 
 To estimate the effect of the off-resonant excitations, we use the fact that the line-width of the transition, whose natural line-width is $\Gamma_D \sim$~1Hz is power broaden such that the FWHM is equal to~\cite{Loudon} $2\sqrt{\Gamma_D^2 + \Omega^2/2} \sim \sqrt{2}\Omega$, since $\Gamma_D << \Omega$ in all cases in this work. This implies that the detuned transition excitation strength scales as $(\Omega_{\Delta m=0}/\Delta)^2$. Also, the following relation holds $\Omega_{\Delta m=0}/\Omega_{\Delta m=-1} \sim 1/(w_0 k)$~\cite{quinteiro2017}. We can then calculate the excitation probability as follows: 
 
 \begin{align}
     p_{\mathrm{pol}}  = \left (\frac{\Omega_{\Delta m=0}}{\Delta}  \right )^2 \times 10^{-2} = \left (\frac{\Omega_{\Delta m=-1}}{\Delta}  \right )^2 \frac{ 10^{-2}}{(w_0k)^2} = \left (\frac{\Omega_0}{\Delta}  \right )^2  \frac{ 10^{-2}}{(w_0k)^2}~S = \left (\frac{\pi}{\tau\Delta}  \right )^2  \frac{ 10^{-2}}{(w_0k)^2}~S =   \frac{ 10^{-6}}{(w_0k)^2}~S
 \end{align}

 where the factor of $10^{-2}$ comes from the error of $1\%$ in the polarization state. 
 
 Additionally, the nearby Zeeman transition, also with $\Delta m=-2$ at $2\pi\times4\,$MHz detuning ($\ket{4^2\text{S}_{1/2}, m=-1/2} \leftrightarrow \ket{3^2\text{D}_{5/2}, m=-5/2}$), is non-dark at the center of the beam due to the presence of the strong electric field gradient~\cite{schmiegelow2012light}, specially since the  magnetic field and $\vec{k}_{729}$ are aligned. This transition could account for spurious excitation either because of the finite pulse length used creates spurious frequency components on this transition or because it is excited off resonantly due to its power broadening. In both cases there is a coefficient of $\frac{\sqrt{5/2}}{(w_0 k)}$ in the Rabi frequencies ratio:  $\Omega_{\Delta m=-2}/\Omega_{\Delta m=-1} = \frac{\sqrt{5/2}}{(w_0 k)}$, where the factor of $\sqrt{5/2}$ comes from Clebsh-Gordan coefficients and the $\frac{1}{(w_0 k)}$ comes from the fact that the $\Delta m=-2$ transition is excited with the transverse gradient~\cite{schmiegelow2012light}.
  
 iii) Considering the Fourier expansion of a square pulse of $\tau=19~\mu$s length (which is the value used throughout this work) the scaling factor turns out to be of $6\times 10^{-4}$ for a $\Delta m=-2$ at $2\pi\times4\,$MHz detuning and therefore the ratio of saturation parameters  between the GSD transition and this detuned carrier is $S_{\tau}/S = \frac{5}{2}\frac{1}{(w_0 k)^2} 6
\times 10^{-4} \approx \frac{10^{-3}}{(w_0 k)^2}$ and the excitation probability reads
 
 \begin{align}
   p_{\tau} =   \sin^2(\sqrt{S_{\tau}}~ \pi/2) \sim (\sqrt{S_{\tau}}~ \pi/2)^2 = \frac{\pi^2}{4}\frac{10^{-3}}{(w_0 k)^2}~S
 \end{align}

\begin{figure}[h]
\centering
\includegraphics[width=0.45\textwidth]{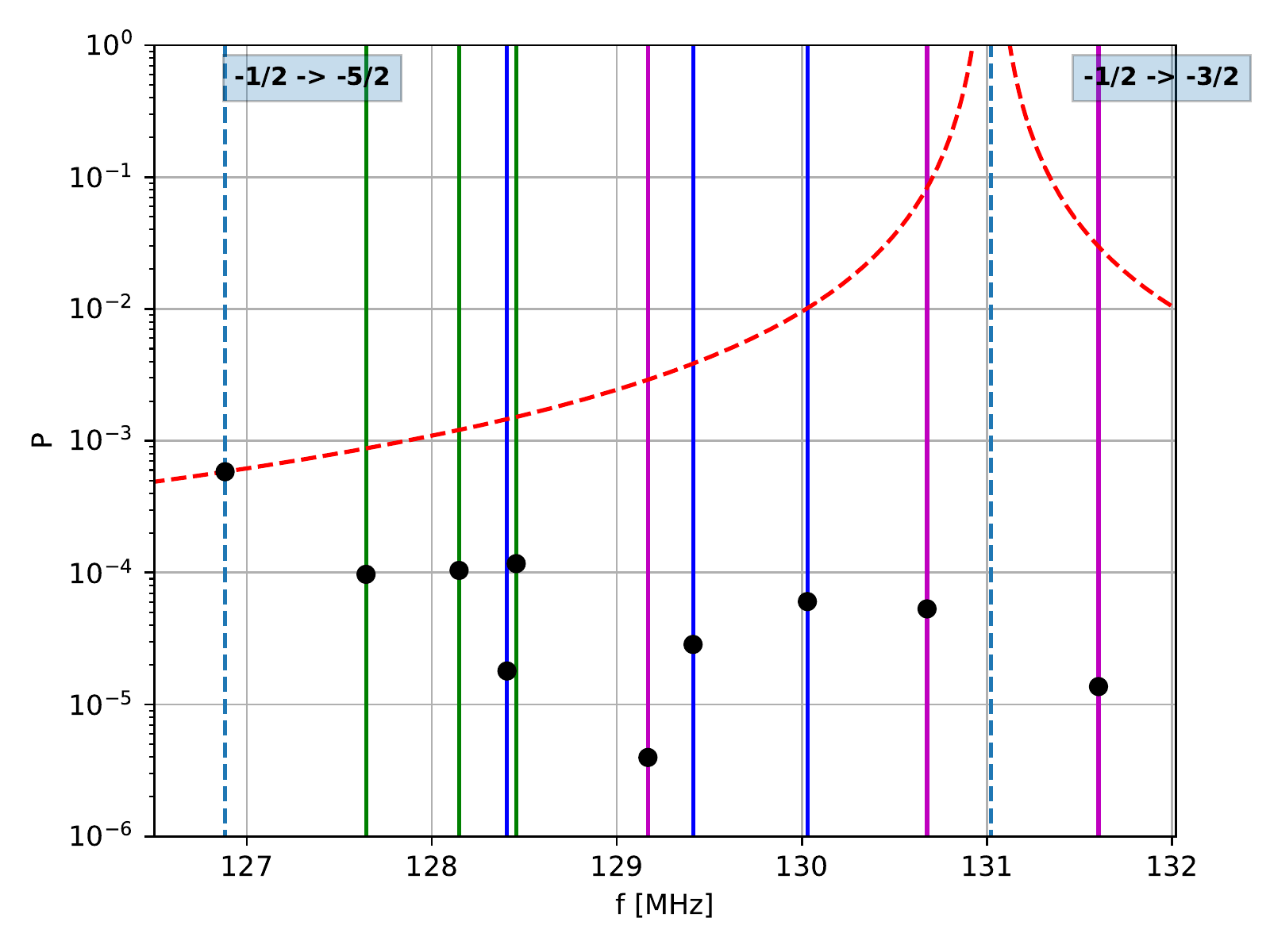}
\caption{\textbf{Errors due to finite pulse length. } One can see that the main contribution comes from the carrier and none of its blue side-bands. }
\label{fig:pulse_length}
\end{figure}

 One could ask whether side-bands of this distant carrier could also be a problem. It is not, as can be seen in the figure \ref{fig:pulse_length}. 
 
The effect of the off-resonant excitations, can be done as before:

\begin{align}
    p_{\Delta} = \left (\frac{\Omega_{\Delta m=-2}}{\Delta}  \right )^2 = \left (\frac{\sqrt{5/2}}{w_0 k} ~ \frac{\Omega_{\Delta m=-1}}{\Delta}  \right )^2 = \frac{5/2}{(w_0 k)^2} ~ \left (\frac{\Omega_0}{\Delta}  \right )^2 S = \frac{10^{-4}}{(w_0 k)^2}~S
\end{align}

Setting a maximum value for $p=0.01$ it is now straightforward to calculate the maximum saturation $S_{\mathrm{lim}}$ that each of these spurious transitions allows in each case. Given $S_{\mathrm{lim}}$ it is possible to calculate the achievable resolution $\sigma_{\text{ePSF}}^{\mathrm{lim}}$ using Eq.1 from the main text, by inserting a power $P = S_{\mathrm{lim}}P_{\mathrm{NS}}$. In this way, Table I from the main text was calculated. 

In all cases, the limit to the saturation in the depletion transition is given by an expression of the form $S_{\mathrm{lim}} = \sqrt{\xi}(w_0 k)^2$, where the dimensionless quantity $\xi$ depends on different experimental parameters other than the beam waist (magnetic field alignment, polarization, pulse time and detunings). Also, the value $P_\mathrm{NS}$ can be exactly calculated in the following way: it is the power that produces a Rabi frequency $\Omega$ at the radius of maximum intensity of the first order Laguerre-Gauss beam (at $r = w_0/\sqrt{2}$) such that $\Omega = \Omega_0 = \pi/\tau$, which results in the following expression for $P_{\mathrm{NS}}$: 

\begin{align}
    P_{\mathrm{NS}} = \frac{2\pi^5 e~ \hbar c}{3\lambda^3\Gamma}~\frac{w_0^2}{\tau^2}
\end{align}

and now we can calculate $\sigma_{\text{ePSF}}^{\mathrm{lim}}$ as follows:

\begin{align}
    \sigma_{\text{ePSF}}^{\mathrm{lim}} = \sqrt{\frac{\pi^2 \hbar c}{3 \lambda^3 \Gamma_\text{D}}} \, \frac{w_0^2}{\tau \sqrt{P_{\mathrm{NS}}S_{\mathrm{lim}}}} = \frac{\lambda}{2\pi\sqrt{2\pi^3e} \sqrt{\xi}}
\end{align}

Remarkably, the achievable resolution does not depend on the beam waist, but on how well the experimental parameters can be controlled to increase $\xi$.

Additionally to fundamental effects, thermal drifts of the optical setup and the piezo stage could limit the achievable resolution. These drifts were determined by checking the position of the beam with respect to the ions every time a measurement concluded, and we observed a drift of $\ll 1~\mu$m per hour. Therefore, this drifts where always happening in time scales longer than the typical duration of our measurements. Thus, drifts over the data aquisition time are much smaller than the achieved resolution. 
Another technical limitation could be stray light scattered off the vacuum windows and a residual component from the diffraction at the fork-grating generating the structured beam. Assuming an ideal Gaussian-shaped beam, we estimate a lower bound to such scattering light power to be $\sim 10^{15}$ times smaller than the total power, thus not affecting $\sigma_\text{ePSF}$.

\section{Numerical calculation of the GSD profiles}

To theoretically calculate the GSD profiles, one needs to calculate the excitation probability $P_D$ as a function of the ion position with respect to the depletion beam $(x_B, y_B)$ and perform a 3-dimensional convolution with the ion's wave packet $|\psi(x_t, y_t, z_t)|^2$. Note that since the wave-packet size is much smaller than the beam waist (and therefore it is also much smaller than the Rayleigh range), we can assume that the beam's transverse intensity profile remains constant. $P_D$ can be calculated as:

\begin{align}
    P_D = \frac{1}{2}\left [ 1-\frac{\cos{(2\Omega \tau)} + 2\Omega\tau \beta \sin{(2\Omega\tau)}}{1 + (2\Omega\tau \beta)^2} \right ]
\end{align}

where $\beta = \eta_{x_t} \bar{n}_{x_t} + \eta_{y_t} \bar{n}_{y_t} + \eta_{z_t} \bar{n}_{z_t}$. $\eta_i = k_{729}\sqrt{\frac{\hbar}{2m\omega_i}}$ are the Lamb-Dicke parameters and  $\bar{n}_i$ the mean phonon number in each direction. Note that the Rabi frequency depends on the coordinates $(x_B, y_B)$ and on the depletion beam waist $w_0$ and total power $P_0$: $\Omega = \Omega(x_B, y_B, w_0, P_0)$. In this way we see how $P_D$ depends on all the relevant parameters as $P_D = P_D(x_B, y_B, w_0, P_0, \tau, \bar{n}_{x_t}, \bar{n}_{y_t}, \bar{n}_{z_t}, \omega_{x_t}, \omega_{y_t}, \omega_{z_t})$. 

For the wave-function $\psi$ we assume that the ion is in thermal equilibrium, and that in every direction the wave-packet is gaussian with a width given by the mean phonon numbers as stated in the main text.

To numerically perform the convolution $P_D * |\psi|^2$ we still need to consider the direction of the beam with respect to the trap geometry (we refer to Figure 1 of the main text). For the fitting of Figure 3 c,d) of the main text, the convolution is made by taking a 3D grid of $512\times512\times512$ points and restricted to a size of $1\times1\times1~\mu m^3$, giving a step size of $2\times2\times2~$nm$^3$.

\section{Exact calculation of the \texorpdfstring{\MakeLowercase{e}}{ePSF}PSF for doughnut and gaussian shaped beams}

In this section we present the exact form of the ePSF for two different beam shapes: one for a first order Laguerre-Gauss as used in the main article, and one for a Gaussian shape. Both cases are done by using a beam waist of $w_0 = 4~\mu$m and a pulse time of $\tau = 20~\mu$s. We present results for low and high beam powers ($1~\mu$W and $120~\mu$W) and we also show the case for two different phonon numbers: $n_{ax} = 1$ and $n_{ax} = 30$. The radial phonon numbers are always kept at a fix value of $n_{rad} = 10$.

\begin{figure}[h]
\centering
\includegraphics[width=0.45\textwidth]{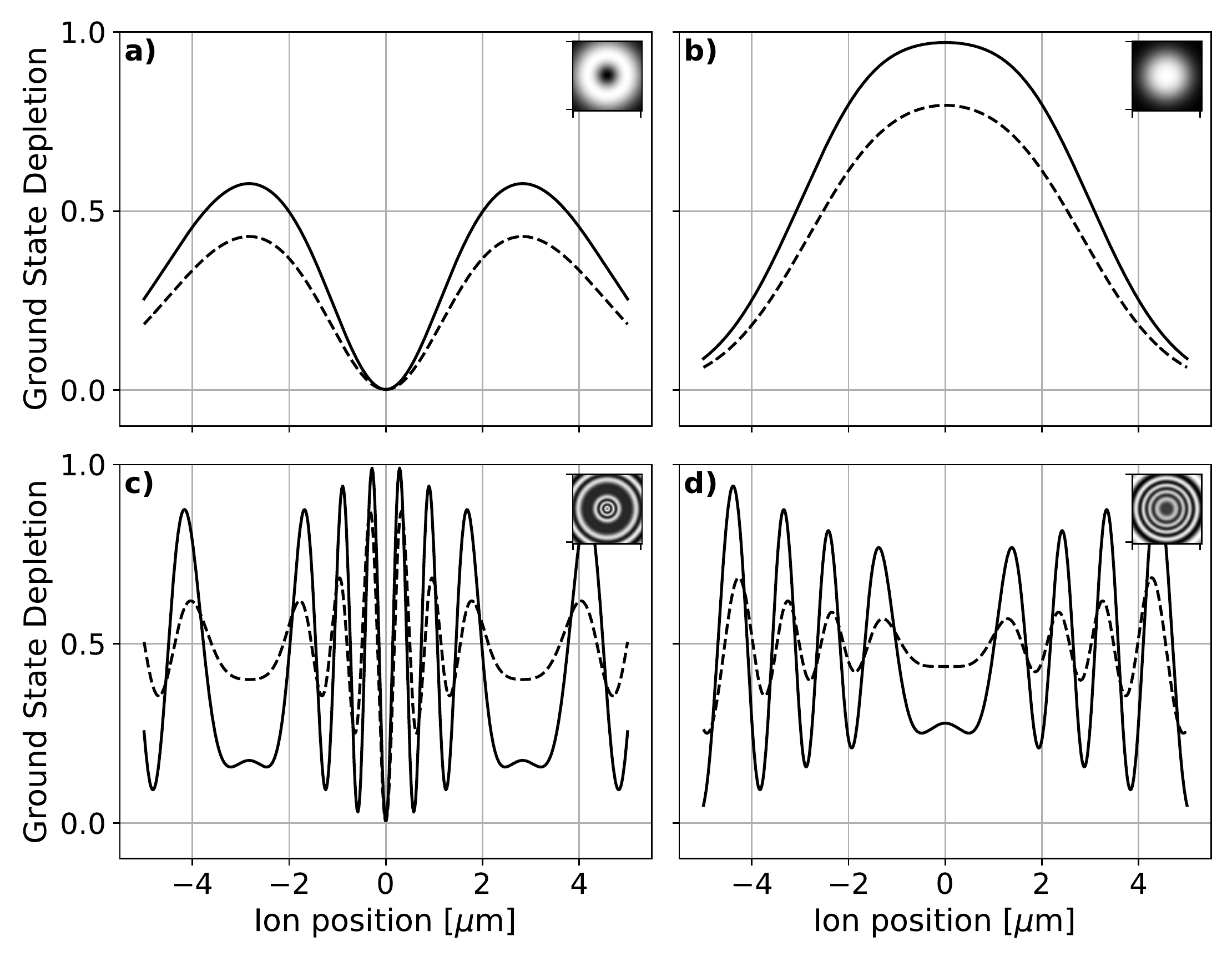}
\caption{\textbf{ePSF with different beam shapes.} Exact calculation of the ePSF for different beam powers, shapes and phonon numbers. In a) and b) the power is $P = 1~\mu$W while in c) and d) it is $P = 120~\mu$W. The beam shape is a first order Laguerre-Gauss in a) and c), and a gaussian in b) and d). The figures show the ePSF for an axial phonon number of $n_{ax} = 1$ (continuous line) and for $n_{ax} = 30$ (dashed line). }
\label{fig:gauss_vs_donut}
\end{figure}

Fig. \ref{fig:gauss_vs_donut} shows how the phonon number (and therefore the ion's temperature) affects the ePSF. The effect vanishes in the very center of the beam in the case of the doughnut shape: this is crucial for the sensing method proposed in the main article, because in the central dark spot, the GSD signal depends only on the ion's wave packet size. 

\bibliography{ions}